\documentclass[twocolumn,notitlepage,prb,color,superscriptaddress,psfig,showpacs,amsmath,amssymb,nobibnotes,longbibliography]{revtex4-2}

\usepackage[colorlinks=true,citecolor=blue]{hyperref}
\usepackage{xspace}
\usepackage{color}
\usepackage{amsmath}
\usepackage{amssymb,stmaryrd}   
\usepackage{mathrsfs}
\usepackage{graphicx,subfigure}     
\usepackage{comment}
\usepackage{indentfirst}

\newcommand{\FFGT}{Fe$_{4}$GeTe$_{2}$\xspace}
\newcommand{\FnGT}{Fe$_{n}$GeTe$_{2}$\xspace}
\newcommand{\FthGT}{Fe$_{3}$GeTe$_{2}$\xspace}

\newcommand{\TC}{$T_\mathrm{C}$\xspace}
\newcommand{\TSR}{$T_\mathrm{SR}$\xspace}
\newcommand{\TQ}{$T_\mathrm{Q}$\xspace}

\newcommand{\HIIab}{$\textbf{H} \parallel ab$\xspace}
\newcommand{\HIIc}{$\textbf{H} \parallel c$\xspace}
\newcommand{\muB}{\mu_\mathrm{B}}
\newcommand{\Ha}{$H_\mathrm{a}$\xspace}
\newcommand{\Hint}{H_{\rm int}}
\newcommand{\HD}{H_{\rm D}}
\newcommand{\Msat}{M_{\rm S}}
\newcommand{\Hres}{H_\mathrm{res}} 
\newcommand{\Hr}{$H_\mathrm{res}$\xspace} 

\newcommand{\FvsH}{$\nu(\Hres)$\xspace}
\newcommand{\Tcross}{T_{\rm cross}}
\newcommand{\Td}{T_{\rm d}}
\newcommand{\Tshape}{T_{\rm shape}}

\definecolor{darkgreen}{rgb}{0, 0.4, 0}

\definecolor{darkbrown}{rgb}{0.4, 0, 0}
\newcommand{\app}[1]{{\color{black} {#1}}}

\definecolor{viol}{rgb}{0.7, 0.4, 1}

\usepackage{placeins}
\usepackage{ulem}
\usepackage{soul}

\usepackage{hyperref}

\definecolor{mygray}{cmyk}{0, 0, 0, 0.3}

\begin{document}

\title{Disentangling the unusual magnetic anisotropy of the near-room-temperature ferromagnet \FFGT}

\author{Riju Pal}
\thanks{Email Address: rijupal07@bose.res.in}
\affiliation{Leibniz Institute for Solid State and Materials Research, Helmholtzstr. 20, D-01069 Dresden, Germany}
\affiliation{Institute for Solid State and Materials Physics, TU Dresden, D-01062 Dresden, Germany}
\affiliation{Department of Condensed Matter and Materials Physics, S. N. Bose National Centre for Basic Sciences, Block JD, Sector III, Salt Lake, Kolkata, 700106, India}

\author{Joyal J. Abraham}
\affiliation{Leibniz Institute for Solid State and Materials Research, Helmholtzstr. 20, D-01069 Dresden, Germany}
\affiliation{Institute for Solid State and Materials Physics, TU Dresden, D-01062 Dresden, Germany}

\author{Alexander Mistonov}
\affiliation{Institute for Solid State and Materials Physics, TU Dresden, D-01062 Dresden, Germany}

\author{Swarnamayee Mishra}
\affiliation{Institute for Solid State and Materials Physics, TU Dresden, D-01062 Dresden, Germany}

\author{Nina Stilkerich}
\affiliation{Institute for Solid State and Materials Physics, TU Dresden, D-01062 Dresden, Germany}
\affiliation{Max Planck Institute for Chemical Physics of Solids, D-01187 Dresden, Germany}

\author{Suchanda Mondal}
\affiliation{Saha Institute of Nuclear Physics, HBNI, 1/AF Bidhannagar, Calcutta 700064, India}

\author{Prabhat Mandal}
\affiliation{Department of Condensed Matter and Materials Physics, S. N. Bose National Centre for Basic Sciences, Block JD, Sector III, Salt Lake, Kolkata, 700106, India}

\author{Atindra Nath Pal}
\affiliation{Department of Condensed Matter and Materials Physics, S. N. Bose National Centre for Basic Sciences, Block JD, Sector III, Salt Lake, Kolkata, 700106, India}

\author{Jochen Geck}
\affiliation{Institute for Solid State and Materials Physics, TU Dresden, D-01062 Dresden, Germany}
\affiliation{Würzburg-Dresden Cluster of Excellence ct.qmat, TU Dresden, 01062, Dresden, Germany}

\author{Bernd Büchner}
\affiliation{Leibniz Institute for Solid State and Materials Research, Helmholtzstr. 20, D-01069 Dresden, Germany}
\affiliation{Institute for Solid State and Materials Physics, TU Dresden, D-01062 Dresden, Germany}
\affiliation{Würzburg-Dresden Cluster of Excellence ct.qmat, TU Dresden, 01062, Dresden, Germany}

\author{Vladislav Kataev}
\affiliation{Leibniz Institute for Solid State and Materials Research, Helmholtzstr. 20, D-01069 Dresden, Germany}

\author{Alexey Alfonsov}
\thanks{Email Address: a.alfonsov@ifw-dresden.de}
\affiliation{Leibniz Institute for Solid State and Materials Research, Helmholtzstr. 20, D-01069 Dresden, Germany}

\date{\today}

\begin{abstract}
In the quest for two-dimensional conducting materials with high ferromagnetic ordering temperature the new family of the layered \FnGT\ compounds, especially the near-room-temperature ferromagnet \FFGT, receives a significant attention. \FFGT features a peculiar spin reorientation transition at \TSR$\sim 110$\,K suggesting a non-trivial temperature evolution of the magnetic anisotropy (MA) -- one of the main contributors to the stabilization of the magnetic order in the low-D systems. An electron spin resonance (ESR) spectroscopic study reported here provides quantitative insights into the unusual magnetic anisotropy of \FFGT. At high temperatures the total MA is mostly given by the demagnetization effect with a small contribution of the counteracting \textit{intrinsic} magnetic anisotropy of an \textit{easy-axis} type, whose growth below a characteristic temperature $\Tshape \sim 150$\,K renders the sample seemingly isotropic at \TSR. Below one further temperature $\Td \sim 50$\,K the \textit{intrinsic} MA becomes even more complex. Importantly, all the characteristic temperatures found in the ESR experiment match those observed in transport measurements, suggesting an inherent coupling between magnetic and electronic degrees of freedom in \FFGT. This finding together with the observed signatures of the intrinsic two-dimensionality should facilitate optimization routes for the use of \FFGT in the magneto-electronic devices, potentially even in the monolayer limit.

\end{abstract}

\maketitle

\section{Introduction}

Magnetic van der Waals (vdW) compounds have become increasingly attractive for fundamental investigations in the past few years since they provide an immense possibility to study quantum magnetic phenomena in low dimensions \cite{burch2018, huang2017, Cai2019}. Furthermore, they appear as very promising functional materials which enable to employ their intrinsic two-dimensionality (2D) for continuous scaling of electronic devices towards the quantum limit \cite{Gong2019_rew}. As such, these layered materials become advantageous candidates for the use in next-generation spintronic devices \cite{khan2020, Li2019, scheunert2016, jimenez2020}. One of their most important parameters from both fundamental and functional points of view is the magnetic anisotropy (MA). First, MA in the presence of a significant intra-layer exchange interaction between magnetic ions enables higher magnetic ordering temperatures (\TC). This is because MA stabilizes long-range magnetic order in low-dimensions, which otherwise would be destroyed at any finite temperature by spin fluctuations according to the  Mermin-Wagner theorem \cite{MerminWagner}. Second, the type and the strength of MA of the dedicated vdW material defines the specific functionalities of the potential spintronic device where this material is used, e.g., it affects the spin-polarization of the generated current.

\begin{figure}[t]
	\centering
	\includegraphics[width=1\linewidth]{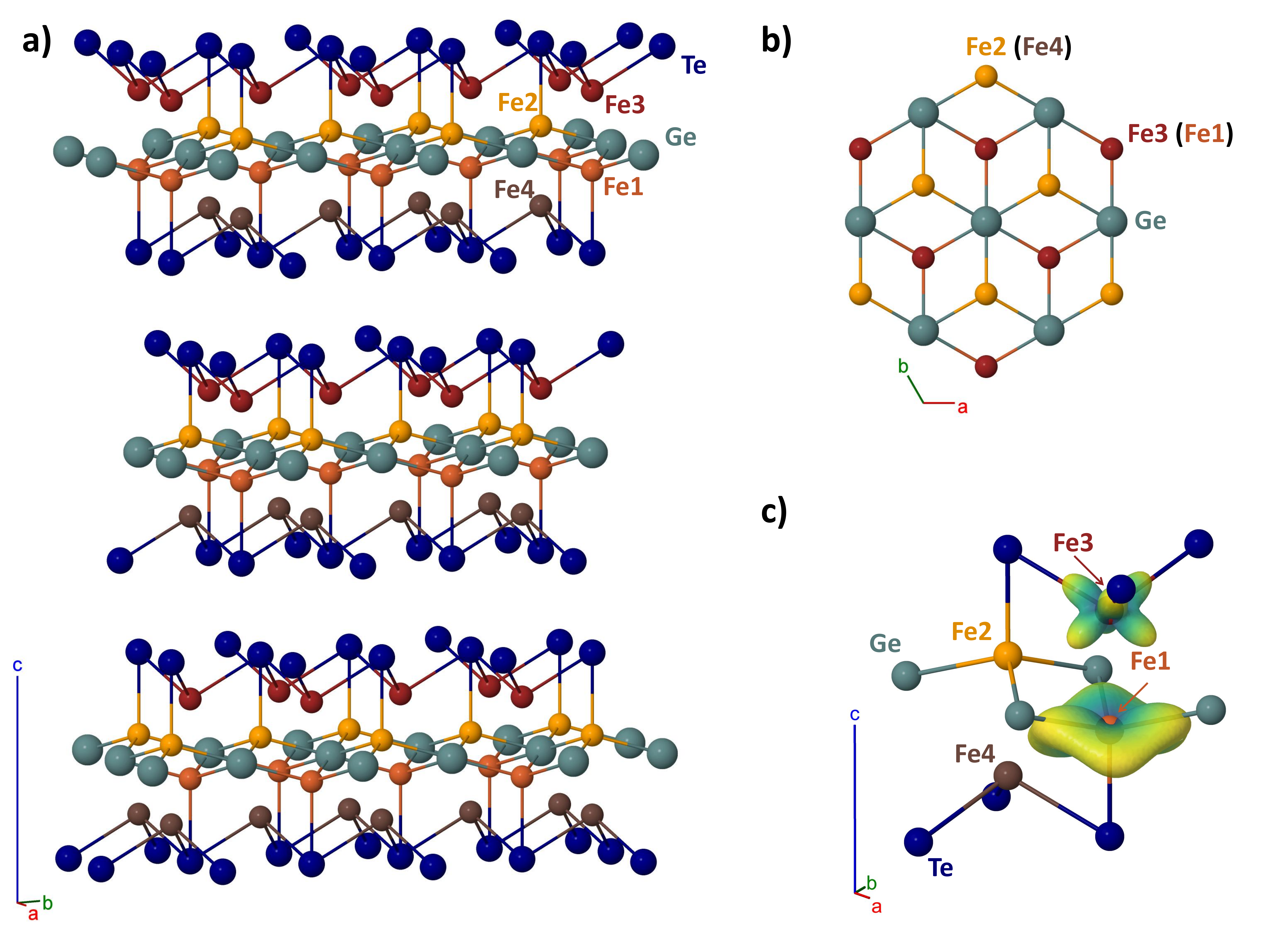}
	\caption{a) Crystal structure of \FFGT featuring van der Waals layers along $c$ direction. b) Top view at the crystal structure of \FFGT. Te atoms are hidden. c) Local crystal structure of four Fe sites. The isosurface represents the crystal field potential, with the size proportional to the overall potential strength. }
	\label{fig:Fig_struc}
\end{figure}

A particularly interesting subclass of the layered magnetic vdW compounds, which currently are promising candidates to fulfill the requirements of high \TC, significant MA and large magnetization at elevated temperatures, is represented by the Fe-rich ferromagnetic \FnGT ($n = 3, 4, 5$) family, where the variation of the Fe content strongly impacts their properties \cite{Chen2013, Liu2017, Fei2018, May2019, May2019asc, Zhang2020, Li2020, Kim2021, Alahmed2021, Ni2021, Bera2023a}. One member of this family, \FFGT \cite{Seo2020}, attracts special attention because of its high, nearly room-temprature value of $\text{\TC} \approx 270$\,K, which likely owes an enhanced interaction between Fe spins arranged according to Fig.~\ref{fig:Fig_struc}(a),(b). Interestingly, the \TC\ of \FFGT is reported to be sensitive to the thickness of the sample and to the Fe concentration \cite{Wang2023}, offering a possibility to further tune the properties of this ferromagnet. Moreover, this compound is capable of generating a very high transport spin polarization \cite{Rana2023}, shows large anomalous Hall conductivity \cite{Pal_arxiv2023, Bera2023prb} and is predicted to demonstrate high performance in a magnetic tunnel junction device \cite{Wu2023}, as well as to generate strong N\'eel spin currents \cite{Shao2023} after doping induced change of the ferromagnetic ground state to an antiferromagnetic one \cite{Seo2021}. The magnetic anisotropy of \FFGT is found to be temperature-dependent with a peculiar change of the preferable spin orientation from in-plane to out-of-plane at the spin reorientation transition temperature \TSR $\sim$ 110 K \cite{Seo2020, Wang2023, Pal_arxiv2023, Bera2023prb, Mondal2021, Bera2023}. Such a change of the spins orientation was explained by the competition between the magnetocrystalline (intrinsic) anisotropy and the shape anisotropy \cite{Seo2020, Wang2023}, however the disentanglement and the quantification of both contributions essential for the understanding of the nature of the peculiar spin reorientation is lacking so far.

To address this intriguing aspect of magnetism of the quasi-2D van der Waals ferromagnet \FFGT  we have performed a detailed electron spin resonance (ESR) spectroscopic study of single crystals of this compound in a broad range of excitation frequencies, magnetic fields and temperatures. In using the ESR spectroscopic technique we exploited its extreme sensitivity to magnetic anisotropies which enables a straightforward measurement and accurate quantification of the MA parameters of the studied magnetic material. Moreover, the direct measurement of the magnetic excitations allowed us to address the dynamic aspects of the magnetism in \FFGT. For instance, we observed that at temperatures up to 300\,K, which is $\sim 30$\,K above \TC, the anisotropy field measured on the time scale of the ESR experiment does not vanish likely due to the sizable short range correlations characteristic of a 2D-spin system. Such an intrinsic low-dimensionality should ensure the persistence of the peculiar magnetic properties when reaching the monolayer limit of the \FFGT sample. As for the ordered state, we found that the evolution of the total magnetic anisotropy field \Ha features several characteristic temperatures. Importantly, the main contribution to the \textit{intrinsic} magnetic anisotropy disentangled in our ESR study is of the uniaxial \textit{easy-axis} type in the entire experimental temperature range. Its growth with decreasing temperature counteracts the demagnetization effect given by the plate-like shape of the sample, which eventually renders the sample seemingly isotropic by approaching the spin reorientation transition temperature \TSR. Moreover, the characteristic temperatures found in the ESR experiment coincide with those observed in the transport measurements \cite{Pal_arxiv2023}, suggesting an intimate interplay between the static and dynamic magnetism, on one side, and the electronic properties of \FFGT, on the other side. Such an interplay appears to provide a new attractive functionality of this material for its application in the next generation spintronic devices, where the transport properties of \FFGT could be tuned by magnetic excitations, and vice versa.

\section{Results}
\label{sec:results}

\begin{figure*}[ht!]
	\centering
	\includegraphics[width=\textwidth]{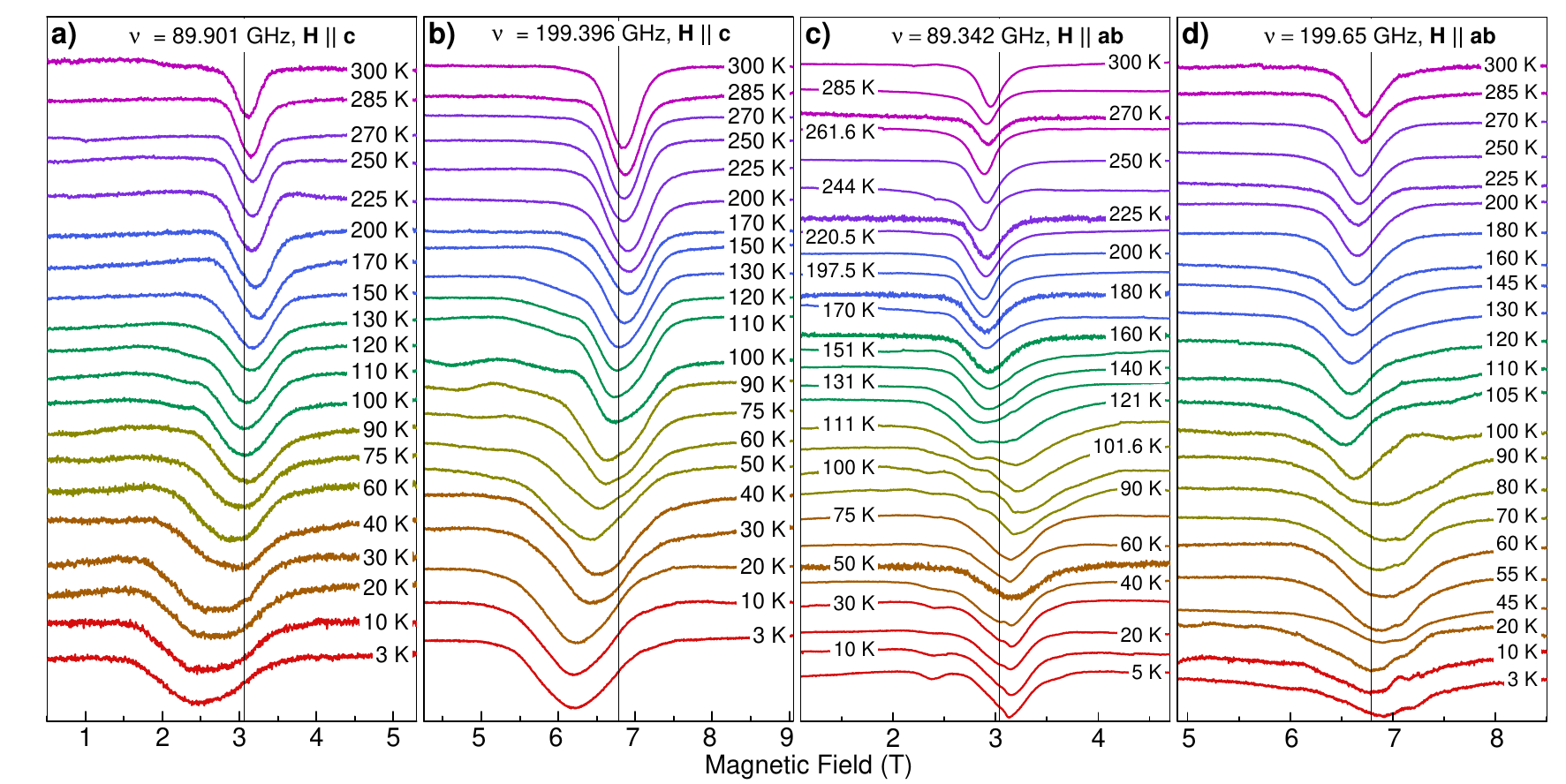}
	\caption{Temperature dependence of the HF-ESR spectra in the \HIIc configuration (a, b) and \HIIab configuration (c, d) at fixed excitation frequencies $\nu \sim 89$\,GHz (a, c) and $\nu \sim 199$\,GHz (b, d). The spectra are normalized to unity and shifted vertically for clarity.  Vertical solid lines represent the expected resonance position of the paramagnetic response according to Eq.~(\ref{eq:Hnu}).}
	\label{fig:Fig_Spectra}
\end{figure*}

For all magnetic field orientations, the ESR spectra of \FFGT consist of a single resonance line, having shoulders at certain temperatures. Representative spectra (the absorption part of the signal $S_{\rm abs}$, see Sec.~\ref{sec:ExptDetails}) recorded at different temperatures, frequencies and field geometries are shown in Fig.~\ref{fig:Fig_Spectra} \app{(also in Figs.~\ref{fig:Fig_Spectra_HIIc}, \ref{fig:Fig_Spectra_HIIab} in Appendix)}. The resonance field \Hr was defined as the position of the minimum of the absorption signal $S_{\rm abs}(H)$.

Vertical solid lines in the panels of Fig.~\ref{fig:Fig_Spectra} and horizontal dashed lines in the panels of \app{Fig.~\ref{fig:Fig3} in Appendix} represent the position of the expected isotropic paramagnetic response at the resonance field $\Hres^{para}$ which is calculated using the conventional paramagnetic resonance relation \cite{Abragam2012}:
\begin{equation}
h\nu = g\muB \mu_{0} \Hres^{para} \ .
\label{eq:Hnu}
\end{equation}
Here $\nu$ is the microwave frequency, g is the g-factor, $h$ is the Plank constant, $\muB$ is the Bohr magneton, $\mu_{0}$ is the free space permeability. The g-factor of $\sim 2.1$ is determined from the frequency dependence of the resonance field \FvsH measured at $T = 300\text{\,K} > \text{\TC} \approx 270$\,K as discussed below in Sect.~\ref{sec:Fdep}.

\subsection{Temperature dependence}
\label{sec:Tdep}

For \HIIc configuration, the HF-ESR spectra were measured at four frequencies $\nu$ = 89.901 GHz, 199.396 GHz, 301.801 GHz and 348 GHz (Fig.~\ref{fig:Fig_Spectra}(a),(b), \app{and Fig.~\ref{fig:Fig_Spectra_HIIc} in Appendix}). As can be seen in these figures, at $T = 3$\,K the ESR lines are at the magnetic field lower than the paramagnetic resonance field $\Hres^{para}$ estimated using Eq.~\ref{eq:Hnu}. Temperature increase causes the shift of the line towards higher magnetic fields continuing up to $\sim 200$ K, and above this temperature the line shifts towards lower magnetic fields. At around $\Tcross \sim 110$\,K, the resonance position of the measured ESR signal matches that of the expected paramagnetic line at $\Hres^{para}$. These observations are practically independent from the excitation frequency, as shown in the panels of Fig.~\ref{fig:Fig_Spectra}. For \HIIab configuration, the ESR response was measured at five frequencies $\nu$ = 75.0405 GHz, 89.342 GHz, 199.65 GHz, 299.88 GHz and 348 GHz (Fig.~\ref{fig:Fig_Spectra}(c),(d), \app{and Fig.~\ref{fig:Fig_Spectra_HIIab} in Appendix}). In contrast to the \HIIc configuration, here at all measurement frequencies the low-$T$ ESR lines have higher resonance fields than the estimated position of the paramagnetic line $\Hres^{para}$. With increasing temperatures from 3\,K to $\sim 200$\,K, the ESR lines are shifting progressively towards lower magnetic fields, crossing the corresponding paramagnetic line position around 110\,K. Above $T \sim 200$\,K the shift of the line changes direction. Interestingly, in both magnetic field configurations the ESR lines develop shoulders and generally change shape with decreasing temperature below $\sim 130$\,K. This is especially well seen in the low frequency data of the \HIIab configuration (Fig.~\ref{fig:Fig_Spectra}(c)).

\begin{figure*}[ht!]
	\centering
	\includegraphics[width=\textwidth]{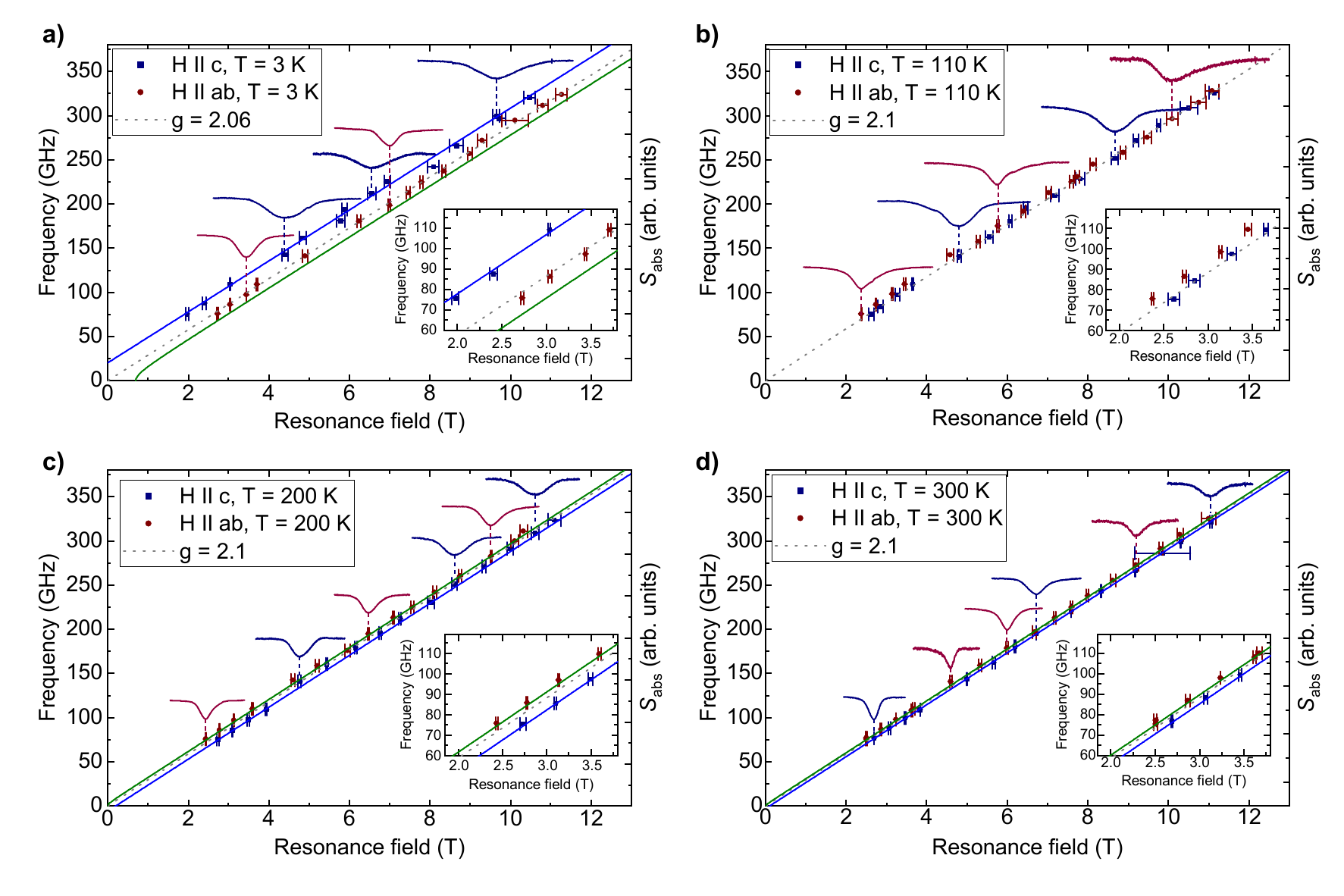}
	\caption{Frequency dependence of the resonance fields \Hr at $T = $ 3\,K (a), 110\,K (b), 200\,K (c), and 300\,K (d), measured in \HIIc (squares) and in \HIIab (circles) configurations, respectively. Dashed lines represent the paramagnetic response according to Eq.~(\ref{eq:Hnu}) with the g-factors given in the corresponding legends. Solid lines depict the results of the best fit using Eqs.~\ref{Eq:EA} and \ref{Eq:EP}. Exemplary, normalized spectra are presented for all temperatures with colors corresponding to the respective magnetic field geometries. Insets: same $\nu-\Hres$ plots zoomed to the low-frequency and low-field regions. }
	\label{fig:Fig4}
\end{figure*}

The quantitative description of the shift of the ESR lines discussed above is presented in \app{Fig.~\ref{fig:Fig3} in Appendix}, where the resonance field \Hr is plotted as a function of temperature at different microwave frequencies and for both \HIIab and \HIIc configurations. There for all frequencies resonance fields measured in different configurations of the magnetic field are crossing around the temperature $\Tcross \sim 110$\,K. The maximum shift of the resonance field from the position of the expected paramagnetic response $\Hres^{para}$ (dashed line) above $\Tcross$ is reached at $\sim 200$\,K. Above this temperature \Hr reduces, however even at $T = 300$\,K, $\sim30$\,K above \TC, the shift of the lines from the expected paramagnetic position remains non zero, as can be also seen in Fig.~\ref{fig:Fig_Spectra}.

\subsection{Frequency dependence}
\label{sec:Fdep}

In order to obtain detailed information on the evolution of the total magnetic anisotropy field \Ha and of the g-factor which both determine the resonance conditions of the ESR response in the FM ordered state, several frequency dependences of the resonance field \FvsH were measured in both configurations of the magnetic field \HIIc and \HIIab, and at a number of selected temperatures $T = 3\text{\,K, \ } 70\text{\,K \ (not shown), \ } 110\text{\,K, \ } 200\text{\,K, \ } 300\text{\,K}$ (Fig.~\ref{fig:Fig4}). To analyze these frequency dependences a linear spin-wave theory with the second quantization formalism \cite{turov, Holstein1940, Alfonsov2021b} was used. The analytical expressions of the spin wave energies for an uniaxial ferromagnet are \cite{turov}:

(a) For easy-axis FM:
\begin{subequations}
	\label{Eq:EA}
	\renewcommand{\theequation}{\theparentequation.\arabic{equation}}
	\begin{align}
	&\text{\HIIc: \ }  & h \nu &= \text{g} \muB \mu_{0} (H+|H_a|)  \label{Eq:EAHIIc}\\
	&\text{\HIIab: \ } & h \nu &= \text{g} \muB \mu_{0} \sqrt{H(H-|H_a|)}  \label{Eq:EAHIIab}
	\end{align}
\end{subequations}

(b) For easy-plane FM:
\begin{subequations}
	\label{Eq:EP}
	\renewcommand{\theequation}{\theparentequation.\arabic{equation}}
	\begin{align}
	& \text{\HIIc: \ }  & h \nu &= \text{g} \muB \mu_{0} (H-|H_a|) \label{Eq:EPHIIc}\\
	& \text{\HIIab: \ } & h \nu &= \text{g} \muB \mu_{0} \sqrt{H(H+|H_a|)} \label{Eq:EPHIIab}
	\end{align}
\end{subequations}
Here the sign of the total magnetic anisotropy field \Ha defines the type of the anisotropy: It is positive for the easy-plane and is negative for the easy-axis anisotropy, respectively. All \FvsH dependences were fitted using these Eqs.~\ref{Eq:EA} and \ref{Eq:EP}. The best fits are presented as solid lines in Fig.~\ref{fig:Fig4}. The results are summarized in \app{Table~\ref{tab:table1} in Appendix} and the obtained values of \Ha are plotted in Fig.~\ref{fig:Fig5} as closed squares and closed circles.

At $T = 300$\,K, i.e., $\sim30$\,K above \TC, and at $T = 200$\,K, i.e., $\sim70$\,K below \TC, the best fits were achieved using the \textit{easy-plane} FM model (Eq.~\ref{Eq:EP}). In the former case (Fig.~\ref{fig:Fig4}(d)) the average value of the anisotropy field is $\mu_0 \text{\Ha} \approx 0.12$\, T, whereas in the latter case (Fig.~\ref{fig:Fig4}(c)) the average value of the anisotropy field can be estimated as $\mu_0 \text{\Ha} \approx 0.2$\, T. At a temperature $T = 110$\,K around $\Tcross$ fits using Eqs.~\ref{Eq:EA} and \ref{Eq:EP} within the error bars yield \textit{zero anisotropy field} (Fig.~\ref{fig:Fig4}(b)). Below $\Tcross$ at $T = 70$\,K the average value of \Ha amounts to $\sim -0.06$\,T. Note, that at this temperature the values of the anisotropy field are negative since the best results were obtained with the \textit{easy-axis} FM model. Within the error bars the g-factor at all these temperatures $T = 300$\,K, $200$\,K, $110$\,K and $70$\,K is found to be practically isotropic with the value of $2.1 \pm 0.01$. 
Finally, at the lowest measurement temperature of 3\,K (Fig.~\ref{fig:Fig4}(a)) the best fit was achieved using the \textit{easy-axis} FM model. Interestingly, the resulting values of the anisotropy field for \HIIc and \HIIab (Fig.~\ref{fig:Fig5} \app{and Table~\ref{tab:table1} in Appendix}) are different, even if the error bars are taken into account, which suggest that such simple uniaxial FM model cannot fully describe the observed magnetic field dependences of the spin wave energies in this low-temperature regime. Additionally, the average g-factor value at this temperature of 3\,K is reduced to $2.06 \pm 0.01$.

Similar ESR measurements were carried out for different orientations of the applied magnetic field in the $ab$-plane of the \FFGT crystal yielding as a result no detectable in-plane anisotropy within the error bars \app{(see Appendix~\ref{app:IP})}.

\subsection{Magnetic anisotropy field}
\label{sec:AnisField}

\begin{figure}[ht]
	\centering
	\includegraphics[width=1\linewidth]{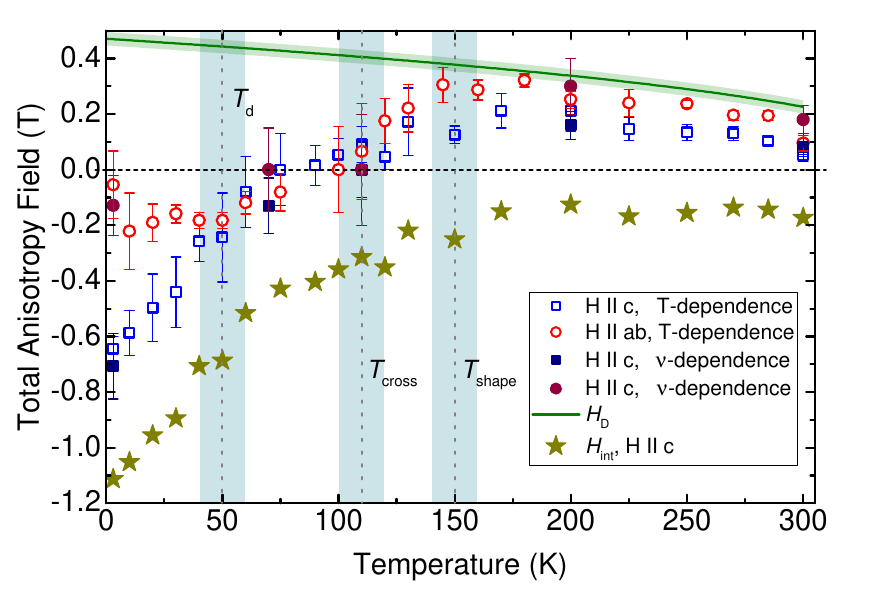}
	\caption{Total anisotropy field \Ha as a function of temperature. Squares represent \Ha for the \HIIc configuration, and circles depict \Ha for the \HIIab configuration obtained from the frequency (closed symbols) and temperature (open symbols) dependences of the resonance position, respectively. Solid line is the calculated anisotropy field $\HD$ given by the shape anisotropy. Stars depict the extracted intrinsic anisotropy field $\Hint$ for the \HIIc configuration. Vertical dashed lines and the shaded area around them indicate characteristic temperatures $\Tshape$, $\Tcross$, and $\Td$ revealed in the analysis of the ESR data, as well as the estimation of the uncertainty region, respectively.}
	\label{fig:Fig5}
\end{figure}

\begin{figure}[ht]
	\centering
	\includegraphics[width=1\linewidth]{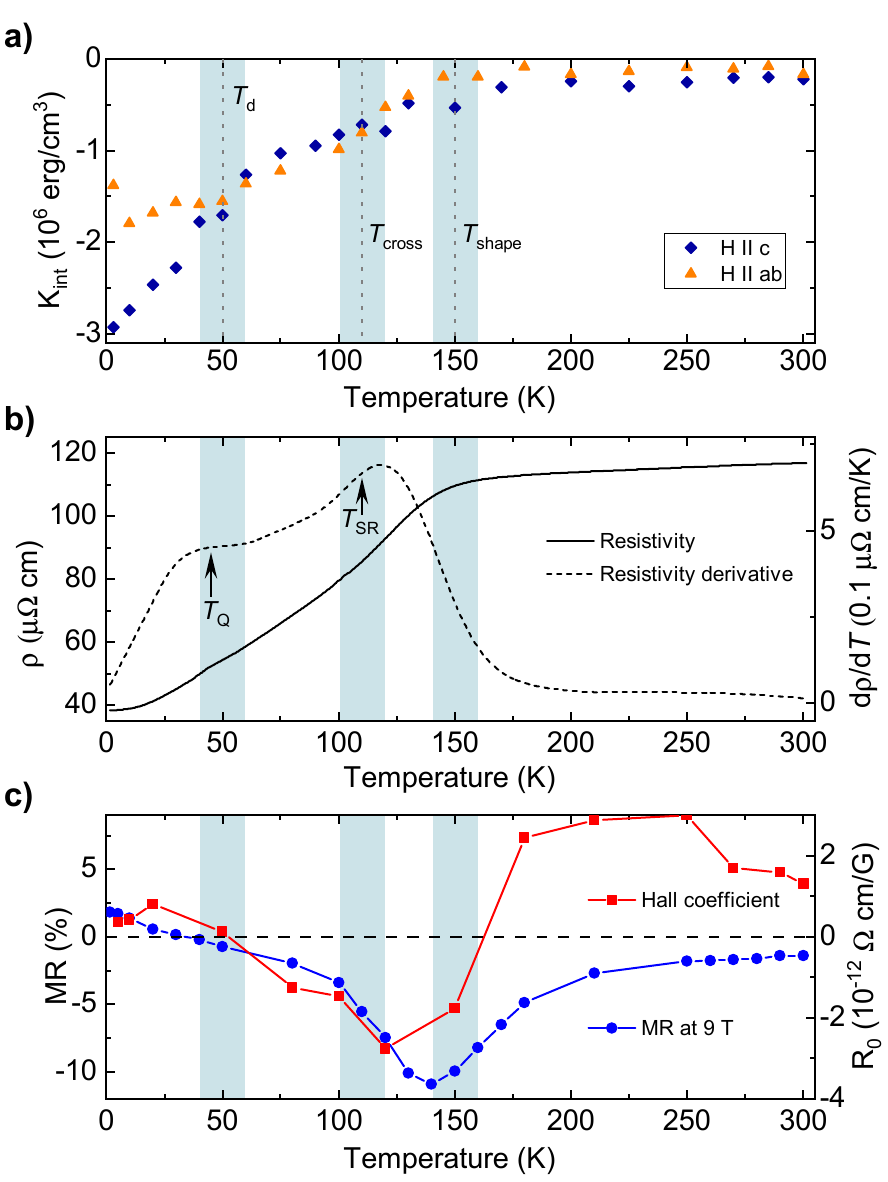}
	\caption{a) Temperature evolution of the intrinsic MA constant $K_{\rm int} = \Hint / (2 \Msat)$ estimated from the measurements in the \HIIc (diamonds) and \HIIab (triangles) configurations, respectively. b) Resistivity and resistivity derivative as a function of temperature. c) Magnetoresistance and ordinary Hall coefficient as a function of temperature. Data in b) and c) is taken from \cite{Pal_arxiv2023}.  Vertical dashed lines and the shaded area around them in a), b) and c) indicate characteristic temperatures $\Tshape$, $\Tcross$, and $\Td$ revealed in the analysis of the ESR data, as well as the estimation of the uncertainty region, respectively.}
	\label{fig:MAE}
\end{figure}

Such a rich set of temperature dependences of \Hr measured at four (for \HIIc) or five (for  \HIIab) frequencies, as shown in Sec.~\ref{sec:Tdep}, enables us to extract the temperature evolution of the total magnetic anisotropy field \Ha by fitting the corresponding \FvsH\ dependences at each temperature using Eqs.~\ref{Eq:EA}, \ref{Eq:EP}. In order to reduce the error bars, during the fitting procedure the g-factor was fixed according to the result of the analysis of the detailed \FvsH dependences with larger amounts of points (Sec.~\ref{sec:Fdep}), and therefore only \Ha\ as the fit parameter was determined using this procedure. At temperatures above $\sim 70$\,K the g-factor was kept constant at a value of 2.1 for both magnetic field configurations \HIIc and \HIIab, and in the temperature range of 3\,K - 70\,K the g-factor was linearly approximated between values of 2.06 and 2.1. The result is shown as open symbols in Fig.~\ref{fig:Fig5}. The good matching of these results with the closed symbols, obtained from the analysis of the detailed \FvsH dependences measured at selected temperatures, validates the use of such approach. 

Interestingly, below $\sim 50$\,K the absolute values of the anisotropy field $|\text{\Ha}|$ obtained in different configurations of the magnetic field, \HIIc (circles) and \HIIab (squares), start to deviate from each other even if the error bars are taken into account: for \HIIc, $|\text{\Ha}|$ continues to increase with decreasing temperature, and for \HIIab, $|\text{\Ha}|$ stays practically constant. First of all this suggests that the model used to describe uniaxial FM spin waves is not fully applicable below $\sim 50$\,K, i.e., the anisotropy becomes more complex than the uniaxial one. Second, from the point of view of the anisotropy, another characteristic temperature $\Td \sim 50$\,K where the spin waves change qualitatively can be identified in addition to the temperature $\Tcross$ where \text{\Ha} changes sign (Fig.~\ref{fig:Fig5}).

The total magnetic anisotropy field entering in Eqs.~(\ref{Eq:EAHIIc}) and (\ref{Eq:EAHIIab}) is the sum of two contributions: $H_{\rm a} = \Hint + \HD$. Here, $\Hint$ is the intrinsic, so-called magnetocrystalline anisotropy field and $\HD = 4 \pi N \Msat$ is the shape anisotropy field determined by the demagnetization factor $N$ of the plate-like sample and its saturation magnetization $\Msat$.
In order to address the question of the origin of the measured total anisotropy field (Fig.~\ref{fig:Fig5}) we have calculated the expected shape anisotropy field with $N \approx 0.71$ estimated from the dimensions of the studied sample \cite{Osborn1945,Cronemeyer1991} (see Sec.~\ref{sec:ExptDetails}), and with $\Msat$ values determined in  \cite{Pal_arxiv2023} (\app{see Appendix~\ref{app:Magn}, Fig.~\ref{fig:Fig_Msat}}). The result is depicted as the  solid line in Fig.~\ref{fig:Fig5}, with the shaded area around it representing the uncertainty of the estimated $\HD$. As can be seen, in the temperature range of 300\,K -- $\sim 150$\,K, the anisotropy field \Ha, obtained from the analysis of the ESR data, has the same sign, temperature evolution and the value quite close to the expected $H_D$. This observation suggests that, at $T > \Tshape \sim 150$\,K, the main contribution to the total magnetic anisotropy of \FFGT is given by the shape of the sample due to the demagnetization effect. Such an estimate of $\HD$ enables to calculate the intrinsic, magnetocrystalline anisotropy field $\Hint$. The result for the \HIIc configuration is shown in Fig.~\ref{fig:Fig5} as stars. As can be seen, it is negative in the whole measurement temperature range, evidencing the \textit{easy-axis} type of the intrinsic magnetocrystalline anisotropy of \FFGT. The respective values of the intrinsic anisotropy constant $K_{\rm int} = \Hint / (2 \Msat)$ for two magnetic field configurations, \HIIc and \HIIab, are presented in Fig.~\ref{fig:MAE}(a). Note, that the values obtained below $\Td \sim 50$\,K should be considered as rough estimates only since, as mentioned above, the simple model used here is not fully applicable to describe the anisotropy at those low temperatures.

\subsection{X-ray diffraction study}
\label{sec:main_XRD}

\begin{figure}[ht!]
	\centering
	\includegraphics[width=1\linewidth]{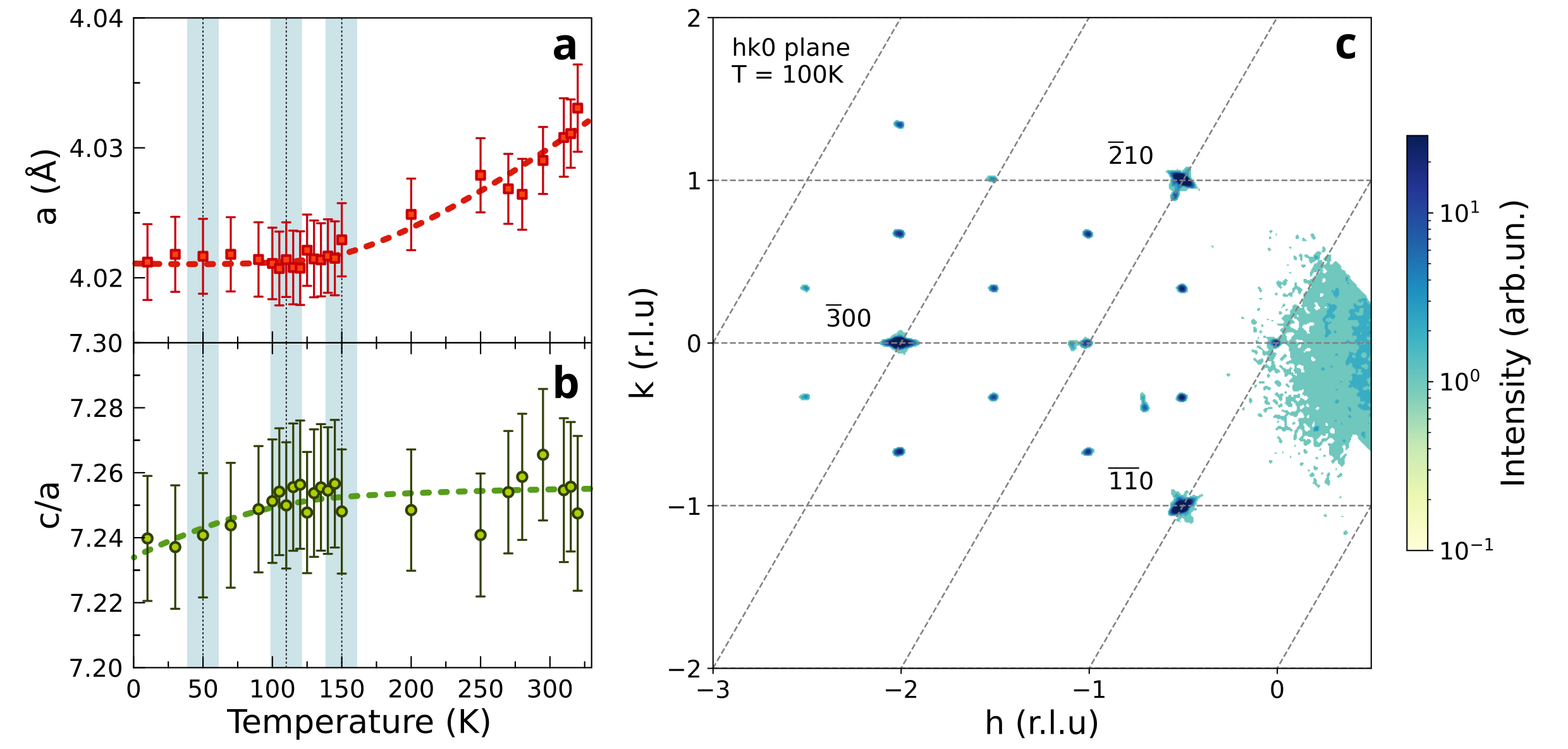}
	\caption{Lattice parameter $a$ (a) and the $c/a$ ratio (b) of the \FFGT as a function of temperature. The dotted lines are guides to the eye. (hk0) cut of the reciprocal space, measured at 100 K (c).}
	\label{fig:structure_byXRD}
\end{figure}

To investigate whether the spin reorientation transition stems from temperature-dependent modifications in the crystal lattice, we conducted single-crystal x-ray diffraction (XRD) measurements across the temperature range of 10~K to 320~K (see Sec.~\ref{sec:ExptDetails} for details). All observed Bragg reflections are fully consistent with the reported $R\bar{3}m$ structure ~\cite{Seo2020}. However, around the Bragg reflections, we discovered strong superlattice reflections with intensities approximately 10\% of the adjacent Bragg reflection (see Fig.~\ref{fig:structure_byXRD}(c)). The positions, intensities, and full width at half maximum (FWHM) of these superlattice peaks displayed no significant variation within the studied temperature range, indicating the stability of the detected superlattice. Notably, the width of the superlattice peaks along the h and k directions was consistently resolution-limited at all studied temperatures, indicating a truly long-range ordered superlattice within the \textit{ab}-plane, with a modulation vector \textbf{q}~=~(1/3, 1/3, 0). Only along the $c$-direction the superlattice is less well correlated, as is often the case with layered materials.

The lattice parameter $a$ and the $c/a$ ratio of the underlying $R\bar{3}m$ structure as a function of temperature are presented in Fig.~\ref{fig:structure_byXRD}(a),(b). As can be observed, both quantities demonstrate a smooth and continuous temperature dependence, implying the absence of any lattice anomaly between 10~K and 320~K in our data. Given the fixed lattice symmetry and the temperature evolution of the lattice constants and the superlattice peaks, we infer that the alterations in the magnetic properties described above are unrelated to a structural transition. Nevertheless, both the parameters $a$ and $c/a$ exhibit minor changes in slope, suggesting the presence of a possible subtle magnetoelastic coupling. It is important to note, however, that the observed effects are minimal and fall below the resolution limit of $\Delta d/d\sim 5\cdot 10^{-3}$ of the current XRD experiment.

\section{Discussion}
\label{sec:Discussion}

One of the central experimental observations of this work is the peculiar temperature evolution of the total anisotropy field in \FFGT (Fig.~\ref{fig:Fig5}). As discussed in Sect.~\ref{sec:AnisField}, at high temperatures the main contribution to the magnetic anisotropy is given by the plate-like shape of the sample due to the demagnetization effect, with a small contribution of the intrinsic \textit{easy-axis} uniaxial anisotropy. Importantly, even at a temperature $\sim 30$\,K above \TC there is a non-zero total anisotropy field, suggesting sizable short range correlations seen on the time scale of the ESR experiment. Such correlations are typical for the intrinsic low-dimensionality of the layered van der Waals materials \cite{Alfonsov2021b, Benner1990, Zeisner2019}, suggesting the 2D-like character of the magnetism in \FFGT. This observation is in line with the results of the analysis of the critical behavior of magnetization isotherms of \FFGT, reported in \cite{Mondal2021}, suggesting a reduced dimensionality ($d<3$) of this compound. Such signatures of the low-dimensional behavior pronounced in the bulk samples suggest that the peculiar magnetic properties observed in this and other works should remain also if the sample is thinned to the monolayer limit, providing an opportunity to utilize \FFGT as a building block of the future nanoscale devices.

Below the characteristic temperature $\Tshape \sim 150$\,K our data reveals a significant growth of the easy-axis anisotropy. Interestingly, below this temperature the electrical resistivity of \FFGT strongly reduces \cite{Seo2020, Wang2023, Pal_arxiv2023, Bera2023prb} (Fig.~\ref{fig:MAE}(b)). Moreover, around $\Tshape$ there is an anomaly in the magnetoresistance and a change of the sign of the Hall coefficient (Fig.~\ref{fig:MAE}(c)). All these observations provide first clues of a plausible relation of the magnetic anisotropy with the properties of the charge carriers in this system. The growing intrinsic \textit{easy-axis} anisotropy with the temperature decrease competes with an \textit{easy-plane} type of the shape anisotropy, making the material seemingly isotropic at around $\Tcross$. This temperature corresponds to the spin reorientation transition at \TSR observed in the static magnetometry and magnetoresistance measurements previously reported in Refs.~\cite{Seo2020, Wang2023, Pal_arxiv2023, Bera2023prb, Mondal2021, Bera2023}. It is important to emphasize, that this spin reorientation effect turns out to be {\it not} an intrinsic property of \FFGT\ but is rather due to the sample plate-like shape of the single crystals. If the sample would be a sphere for which $H_D$ is zero, then the total anisotropy would be always negative (\textit{easy-axis}) in the measurement temperature range and never positive (\textit{easy-plane}). Our experimental result stands in contrast with theoretical calculations of the magnetocrystalline anisotropy for the monolayer \cite{Kim2021, Yang2021, Kim2022} or for a bulk  sample of \FFGT \cite{Kim2022,Ghosh2023}, which predict the anisotropy to be of the \textit{easy-plane} type. However, the type of the uniaxial anisotropy calculated for the bulk sample in \cite{Yang2021} and for the monolayer sample using GGA+U approach in \cite{Ghosh2023} is in agreement with our experimental data.

Below one further characteristic temperature $\Td \sim 50$\,K the anisotropy becomes more sophisticated than a simple uniaxial easy-axis type. Remarkably, this characteristic temperature is also related to the observations in the transport experiments, where around temperature \TQ $\sim 40\text{\,K} - 50\text{\,K}$ the change of the sign of the magnetoresistance and Hall coefficient \cite{Pal_arxiv2023} takes place together with the occurrence of an anomaly in the first derivative of the resistivity \cite{Seo2020, Pal_arxiv2023, Bera2023prb} (Fig.~\ref{fig:MAE}(b),(c)). Moreover, as it is especially well seen in the temperature dependence of the ESR spectra measured in the \HIIab configuration at low frequencies (Fig.~\ref{fig:Fig_Spectra}(c), \app{and Fig.~\ref{fig:Fig_Spectra_HIIab}(a-c) in Appendix}), the ESR signal changes its shape in the temperature range of 3\,K -- $\sim 130$\,K. Starting from low temperatures, the signal develops a shoulder at the low field side, and with increasing $T$ the shoulder grows in intensity, whereas the main line gets smaller. At around $\Tcross$ which is related to \TSR their intensities become almost equal, and above this temperature the former shoulder transforms into the dominant line of the total ESR response. One can conjecture that such a behavior can indicate a gradual, \textit{qualitative} change of the spin wave modes in \FFGT\ contributing to the ESR response as a function of temperature. 
Apparently, these modes coexist in the broad temperature range. Such coexistence is likely due to the complexity of the crystal and electronic structures of \FFGT. The crystal structure features two crystallographically inequivalent Fe ions, which can be grouped into two pairs Fe1/Fe2 and Fe3/Fe4 considering their local environment (Fig.~\ref{fig:Fig_struc}(a),(b)). Also, a contribution of the orbital degree of freedom from the Fe ions as well as nearest ligands Te and Ge, suggested by the enhanced g-factor in comparison with the free electron value may play an important role in the magnetic and electronic properties of \FFGT. Indeed, the X-ray Magnetic Circular Dichroism (XMCD) studies reported in \cite{Kim2022} suggest that, even though the orbital moment is reduced with respect to \FthGT, it is found to be of $\sim 0.05$\,$\muB / {\rm Fe}$ in the \FFGT compound.

In order to address the question of the origin of the magnetic anisotropy in \FFGT responsible for the stabilization of the ground state with the respective spin wave excitations, one has to understand the nature of the magnetism first. Without doubt an important role there is played by the conduction electrons, however an additional contribution of the electrons localized at the Fe $d$-orbitals is suggested by the DFT calculations in \cite{Wang2023, Liu2022} and by the analysis of the critical magnetization isotherms in \cite{Mondal2021}. 

It is instructive to consider first an itinerant picture. The DFT calculations in \cite{Pal_arxiv2023} suggest that there could be two different stable states, separated by a small energy difference of $\sim 10$\,meV $\approx 116$\,K. Increase of the temperature may assist population of the state with higher energy, and therefore induce a change in the spectrum of the excitations probed in the ESR experiment. However, in this case little is known about the details of the anisotropies of these states, only that the preferable orientation of the Fe spins is close to the $c$-axis, which is qualitatively in-line with observations of the ESR line shifts.

In order to address the problem of the anisotropies, it is possible to write down the crystal field (CF) or the ligand field potential, felt by the $d$-electrons of Fe$^{2+}$, for the arbitrary symmetry of the local environment. Such CF potential of the general form can be expressed as:

\begin{align}
V_{CF}  = B_2^0 C_2^0 + B_2^1 (C_2^{-1}-C_2^{+1}) & + B_2^2 (C_2^{-2}+C_2^{+2}) \nonumber \\
+ B_4^0 C_4^0 + B_4^1 (C_4^{-1}-C_4^{+1}) & + B_4^2 (C_4^{-2}+C_4^{+2}) \nonumber \\
+ B_4^3 (C_4^{-3}-C_4^{+3}) & + B_4^4 (C_4^{-4}+C_4^{+4})
\label{Eq:CF}
\end{align}
Here, $B_k^q$ are the crystal field parameters, and $C_k^{q}$ are the spherical tensor operators related to spherical harmonics $C_k^{q} = \sqrt{\frac{4 \pi}{2 k + 1}} Y_k^q$.
With the assumption that all 6 $d$-electrons occupy the Fe$^{2+}$ orbitals (localized picture), it is possible to estimate the coefficients $B_k^q$ describing the local crystal field potential for all four Fe ions using the simple point charge model. Though this approach is applicable mostly to the ionic compounds, as it neglects effects from covalency and the itinerant charge carriers, it can still give a qualitative understanding of the symmetry of the electronic local environment \cite{Barnes1981}. This is important in this compound even if only the itinerant magnetism is considered \cite{Kim2022}. Following the procedure described in Ref.~\cite{Song2022}, the calculated CF potential (see isosurfaces in Fig.~\ref{fig:Fig_struc}(b)) turns out to be the same within each pair Fe1/Fe2 and Fe3/Fe4, however it is quite different between the pairs. The non-equivalence of the Fe sites from the DFT point of view is also suggested in \cite{Li2020, Kim2021, Wang2023, Rana2023, Kim2022, Ghosh2023, Liu2022}. Interestingly, in both cases the sign of the estimated second order term $B_2^0$ suggests an \textit{easy-axis} anisotropy, which is in-line with our experimental observation of the ESR line shift in the whole measurement temperature range. Additionally, the CF potential at the Fe3 (Fe4) position is smaller than that at the Fe1 (Fe2) position, and the contribution of the second order terms $B_2^q$ is smaller than the cubic ones $B_4^q$, which suggests that the magnetic anisotropy might be smaller for the Fe3 (Fe4) ions. Such a difference in the CF potentials for different Fe ions is mostly due to a difference of the charge of the nearest ligands, Ge$^{4-}$ in the case of Fe1 and Fe2, and Te$^{2-}$ in the case of Fe3 and Fe4. Therefore, the presence of two different local Fe environments and potentially non-equivalent exchange interactions coupling the respective Fe spins suggest at least 4 magnetic sublattices with two different types of anisotropy. 

Moreover, considering a complex, unusual behavior of the magnetic anisotropy as a function of temperature revealed in the present ESR study, as well as a number of features in the static magnetization and magnetotransport measurements observed at the characteristic temperatures established in our ESR experiments (Fig.~\ref{fig:MAE}), it is also possible that these peculiarities may be due to the coexistence of itinerant and localized magnetism in \FFGT. In both cases, the local Fe-ligand structure suggests an existence of several distinct (localized or itinerant) magnetic sublattices in \FFGT, with their unique anisotropies and interactions, which contribute to the formation of the FM ordered ground state and the respective excitations. 
Based on our XRD study, both the average $R\bar{3}m$ lattice structure and the superlattice exhibit no discernible anomaly within the investigated temperature range, particularly between 100~K and 150~K. This suggests that the amplification of the intrinsic easy-axis anisotropy is not attributed to a structural transition, however there could be a minor structural contribution due to a $c/a$ parameter, exhibiting in its temperature dependence a possible change in slope. Moreover, one might speculate that the additional potential arising from the superlattice modulation interacts with electronic states, possibly inducing variations in electronic properties as temperature decreases. In principle, this could also impact the parameters $B_k^q$ in Equation \ref{Eq:CF}, thereby influencing the magnetic anisotropy. However, this scenario certainly requires further scrutiny in future studies.
It is interesting, that despite such a complexity of the system, the high temperature excitations at $T > \sim 150$\,K are well described by the simplest one-sublattice ferromagnetic model with the predominant contribution of the shape anisotropy, suggesting a temperature averaging of the differences in the local anisotropies of the individual magnetic sublattices and of the interactions coupling them. 

Nevertheless, a remarkable correspondence of the characteristic temperatures observed in the ESR and transport experiments suggests a plausible strong mutual dependence of the spin waves and the charge carriers. That, in turn, opens a possibility to tune the transport properties of \FFGT by controlling the magnetic excitations, and \textit{vice versa}, which could find an application in the next generation spintronic devices. So far this kind of coupling has been demonstrated mostly on the artificial hybrid ferromagnet/metal multilayers, as, e.g., in Ref.~\cite{Saitoh2006}, whereas \FFGT may be an example of ``natural'' functional material featuring an intrinsic magnetoelectronic coupling. 

\section{Conclusion}

In summary, we have performed a detailed ESR spectroscopic study of the single-crystalline quasi-2D van der Waals  ferromagnet \FFGT with the high ordering temperature $T_{\rm C} = 270$\,K. The measurements were carried out in a broad range of excitation frequencies and temperatures, and at different orientations of the magnetic field with respect to the sample. One of the main observations of this work is the unusual evolution of the magnetic anisotropy field with temperature. First of all, it does not vanish at temperatures well above \TC, likely due to the sizable short range correlations. This suggests an intrinsic low-dimensionality, which in turn should ensure the persistence of the observed magnetic properties when reaching the monolayer limit of \FFGT sample. The analysis of the ESR data reveals several characteristic temperatures for the magnetic behavior of \FFGT in the ferromagnetically ordered state. In between $T_{\rm C}$ and  the characteristic temperature  $\Tshape \sim 150$\,K the total anisotropy is dominated by the demagnetization effect due to the plate-like shape of the sample. However, below $\Tshape$ the intrinsic uniaxial easy-axis magnetic anisotropy starts to noticeably grow counteracting the easy-plane shape anisotropy, which renders the sample seemingly isotropic at the spin reorientation transition at \TSR. The data reveals one further characteristic temperature $\Td \sim 50$\,K, below which the anisotropy becomes considerably more complex, likely due to a multi-sublattice nature of the ground state stabilized in this temperature regime. Importantly, our results give evidence that the main contribution to the intrinsic magnetic anisotropy of \FFGT is always of an \textit{easy-axis} type in the entire temperature range below and above \TSR, and that the reorientation of the sample's magnetization from the out- to the in-plane direction observed at \TSR in the static magnetic measurements results from the competition between the intrinsic easy-axis anisotropy of this compound and the (extrinsic) shape anisotropy of the plate-like crystal. The ESR spectral shape exhibits a complex temperature evolution suggesting  a gradual, \textit{qualitative} change of the spin wave modes in \FFGT. This is likely due to the complexity of the crystal and electronic structures of \FFGT yielding several distinct magnetic sublattices that feature different temperature dependent anisotropies and interactions. Interestingly, our XRD study suggests that the temperature evolution of the anisotropy is not driven by the structural transition. The characteristic temperatures found in the ESR experiment are in a remarkable correspondence  with those observed in the transport measurements, suggesting intertwined magnetic and electronic behaviors in \FFGT. Altogether, our findings on the peculiar intrinsic magnetic anistropy of \FFGT provide important clues for the functionalization of this metallic  near-room-temperature ferromagnet for the use in magneto-electronic devices, and call for the development of the microscopic theories, as well as for the experiments devoted to the detection of the spin Hall and inverse spin Hall effects, for a better understanding of a complex interplay of magnetic and electronic degrees of freedom in this material.


\section{Experimental Section}

\label{sec:ExptDetails}

\subsection{Samples}
Single crystals of \FFGT studied in this work were grown with the standard chemical vapor transport (CVT) method with iodine as a transport agent. The details of the synthesis and comprehensive analytical and physical characterization of these grown crystals proving their high quality were reported in Refs.~\cite{Pal_arxiv2023, Mondal2021}. In particular, they crystallize in a rhombohedral structure with space group $R\Bar{3}m$ and have lattice parameters $a = b = 4.036 \pm 0.003$\,\AA, $c = 29.2 \pm 0.1$ \AA, $\gamma = 120 \pm 0.1$\,$^{\circ}$ (hexagonal representation), close to the previously reported values \cite{Seo2020}. For the electron spin resonance (ESR) study, a bulk crystal of \FFGT with the lateral dimensions of $\sim 1.2$\,mm $\times$ $\sim 0.6$\,mm $\times$ $\sim 0.15$\,mm was chosen. The characteristic temperatures $\text{\TC} \approx 270$\,K and $\text{\TSR} \approx 110$\,K of this particular sample were reconfirmed by the DC magnetization measurements with a magnetic field ($\textbf{H}$) of 100 Oe applied both parallel to the $ab$-plane and the $c$-axis, respectively (See \app{Fig.~\ref{fig:Fig_Msat} in Appendix}). This measurement was performed in a superconducting quantum interference device vibrating sample magnetometer (SQUID-VSM) from Quantum Design. To remove any remanent or residual magnetization present in the sample, an external magnetic field of 5\,T was applied at 300\,K and then gradually reduced to zero while periodically changing the direction of the field's polarity. All the measurements were done in the zero field cooled (ZFC) condition, where the sample was cooled down to 1.8\,K from 300\,K without field and data was taken during warming up with a constant field. 

\subsection{X-ray diffraction}
Single crystal x-ray diffraction (XRD) measurements were performed in 10 –- 320~K temperature range at a high-performance laboratory XRD facility ``VEGA'' (Versatile Extreme conditions Generating diffraction Apparatus), optimized for resolution and sensitivity. This custom-made instrument is equipped with a monochromatized Mo K$_{\alpha}$ radiation source ($\lambda = 0.70930$\,\AA) and a CdTe area detector DECTRIS Pilatus with 300 000 pixels and no readout noise for high detection efficiency and minimum background. The cooling of the sample down to 10~K was accomplished by a low-vibration pulse-tube cryostat, which was itself mounted on a specialized four-circle diffractometer. A detailed study of the 100 -- 150~K range has been performed due to expected structural changes accompanied by spin-reorientation transition~\cite{Pal_arxiv2023}.

Data collection and reduction were performed with the CrysAlisPro package program from Oxford Diffraction \cite{Crysalis}. Based on the indexation of CrysAlisPro, the lattice parameters $a$ and $c$ of the hexagonal unit cell were analyzed using the python library pyfai \cite{Kieffer2020} and a weighted minimization algorithm over 16 peaks. The structure solution is performed by Superflip~\cite{Palatinus2007}, which is employed in JANA2020~\cite{Petricek2023}, and subsequent refinements are performed in the JANA2020 software package. The data was plotted using MagicPlot software~\cite{MP}. The crystal structure shown in Fig.~\ref{fig:Fig_struc} was created using Jmol software \cite{Jmol}.

\subsection{Electron spin resonance}
The ESR measurements were performed using a home-made high-field/high-frequency electron spin resonance (HF-ESR) spectrometer. Here, a vector network analyzer (PNA-X) from Keysight Technologies with the extensions from Virginia Diodes, Inc. (VDI) were used for the generation and detection of microwaves (MW) in the frequency range from 75\,GHz to 330\,GHz. Higher frequencies were generated with a set of amplifier multiplier chains (AMC) and detected by a set of zero bias detectors (ZBD), all from VDI. For continuous magnetic field sweeps, a superconducting magnet from Oxford Instruments was used. The homemade probehead with oversized waveguides was inserted into a variable temperature insert (VTI) of a He$^{4}$ cryostat (from Oxford Instruments) to obtain stable temperatures of the sample from 300\,K down to 3\,K. All measurements were performed by continuously varying the magnetic fields from 0\,T to 16\,T and back at a constant temperature and frequency. The ESR spectra were recorded with  the external magnetic field applied along the $c$-axis and along two orthogonal directions within the $ab$-plane, respectively. 

\subsection{Analysis of the HF-ESR spectra}
In the employed HF-ESR setup based on PNA-X, the instrumental mixing of the absorption $S_{\rm abs}$ and dispersion $S_{\rm disp}$ components of the detected signal $S_\mathrm{D}$ is unavoidable due to the complex impedance of the broad-band probehead. Therefore both components of the measured signal at the detector $S_\mathrm{D}^{\rm amplitude}$ and $S_\mathrm{D}^{\rm phase}$ can be presented as: 	
\begin{align}
S_\mathrm{D}^{\rm amplitude} & \propto S_{\rm abs}\sin (\alpha) + S_{\rm disp}\cos (\alpha) \nonumber\\
S_\mathrm{D}^{\rm phase} & \propto  S_{\rm abs}\sin (\alpha) - S_{\rm disp}\cos (\alpha) \nonumber
\end{align}
where $\alpha$ is a parameter which defines the degree of instrumental mixing of the absorption and dispersion components. However, since the vector network analyzer measures both the amplitude $S_\mathrm{D}^{\rm amplitude}$ and the phase $S_\mathrm{D}^{\rm phase}$ of the transmitted MW radiation simultaneously, the absorption component of the signal $S_{\rm abs}$ can be rectified from the as-measured spectra by varying $\alpha$ parameter:
\begin{align}
S_{\rm abs} & \propto  S_\mathrm{D}^{\rm amplitude}\sin (\alpha) + S_\mathrm{D}^{\rm phase}\cos (\alpha)
\label{eq:Sabs}
\end{align}
In the case of the noisy phase measurement channel, $S_\mathrm{D}^{\rm phase'}$ can be numerically reconstructed applying the Kramers-Kronig transformation to the signal detected in the amplitude measurement channel $S_\mathrm{D}^{\rm amplitude}$. The resulting $S_\mathrm{D}^{\rm phase'}$ is then used in Eq.~\ref{eq:Sabs} to obtain the absorption component of the signal $S_{\rm abs}$. In order to make sure that $S_{\rm abs}$ is reliably rectified both procedures -- direct utilization of the measured $S_\mathrm{D}^{\rm phase}$, or calculation of $S_\mathrm{D}^{\rm phase'}$ -- were used in this work.

\begin{acknowledgments}

This work was supported by the Deutsche Forschungsgemeinschaft (DFG, German Research Foundation) through grants No. KA 1694/12-1, AL 1771/8-1, project ``Nematische Suszeptibilität und Quantenoszillationen in FeSe und LaFePO unter einachsigem Druck'' (project-id 429883427), and within the Collaborative Research Center SFB 1143 ``Correlated Magnetism – From Frustration to Topology'' (project-id 247310070), and the Dresden-Würzburg Cluster of Excellence (EXC 2147) ``ct.qmat - Complexity and Topology in Quantum Matter'' (project-id 390858490). A.N.P. acknowledges financial support from DST-Nano Mission Grant No. DST/NM/TUE/QM-10/2019. R.P. acknowledges DST, SNBNCBS, and IFW for funding. S.M. and J.G. express gratitude to the Dresden-Würzburg Cluster of Excellence on Complexity and Topology in Quantum Matter (ct.qmat EXC-2147, Project No. 390858490) for financial support.
\end{acknowledgments}

\bibliography{References_ESR.bib}


\clearpage

\FloatBarrier

\onecolumngrid

\section{Appendix}


\begin{figure*}[ht!]
	\centering
	\includegraphics[width=\textwidth]{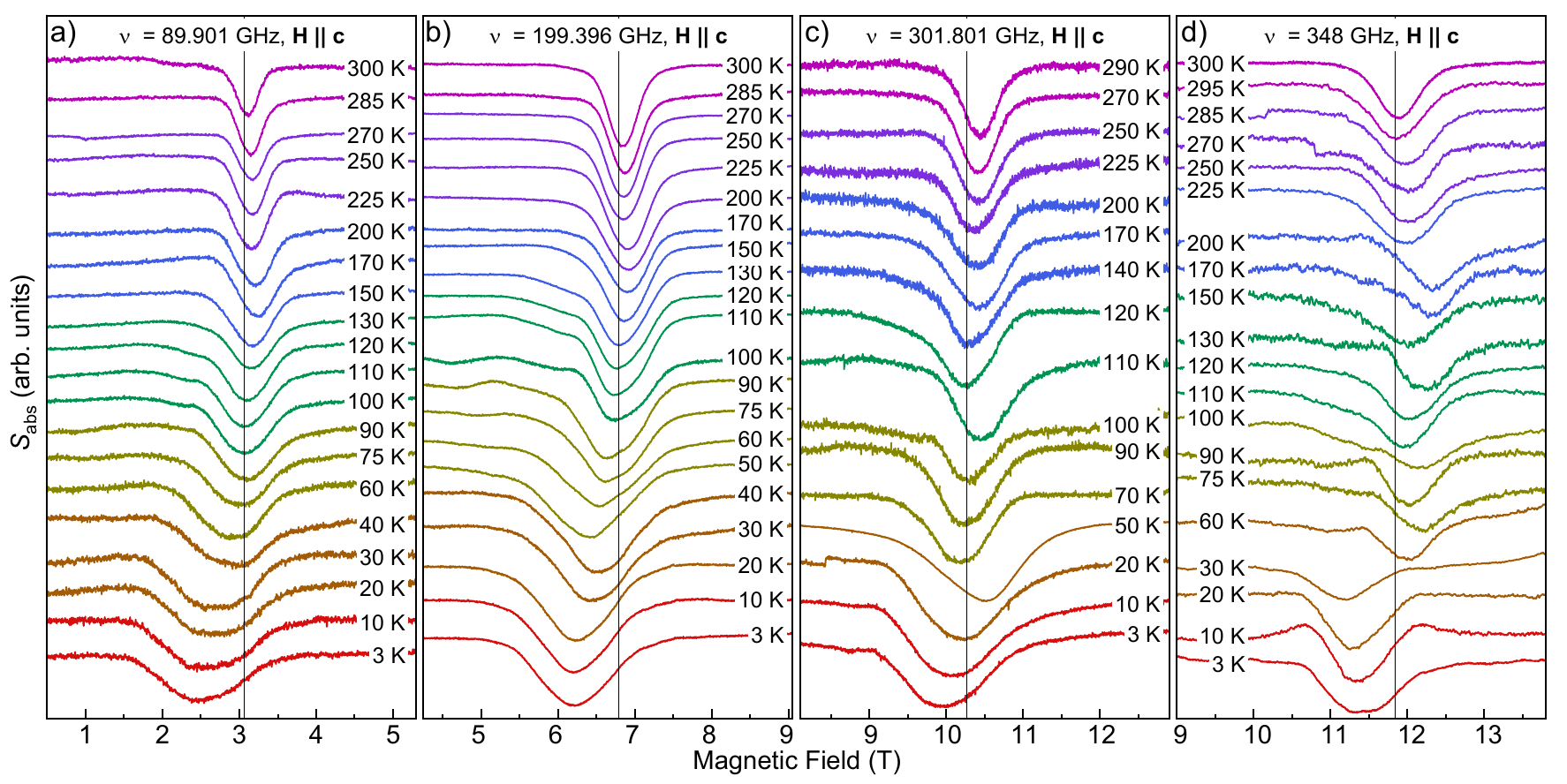}
	\caption{Temperature dependence of the HF-ESR spectra in the \HIIc configuration at fixed excitation frequencies $\nu = $ 89.901\,GHz (a), 199.396\,GHz (b), 301.801\,GHz (c), and 348\,GHz (d). The spectra are normalized to unity and shifted vertically for clarity.  Vertical solid lines represent the expected resonance position of the paramagnetic response.}
	\label{fig:Fig_Spectra_HIIc}
\end{figure*}
\begin{figure*}[ht!]
	\centering
	\includegraphics[width=\textwidth]{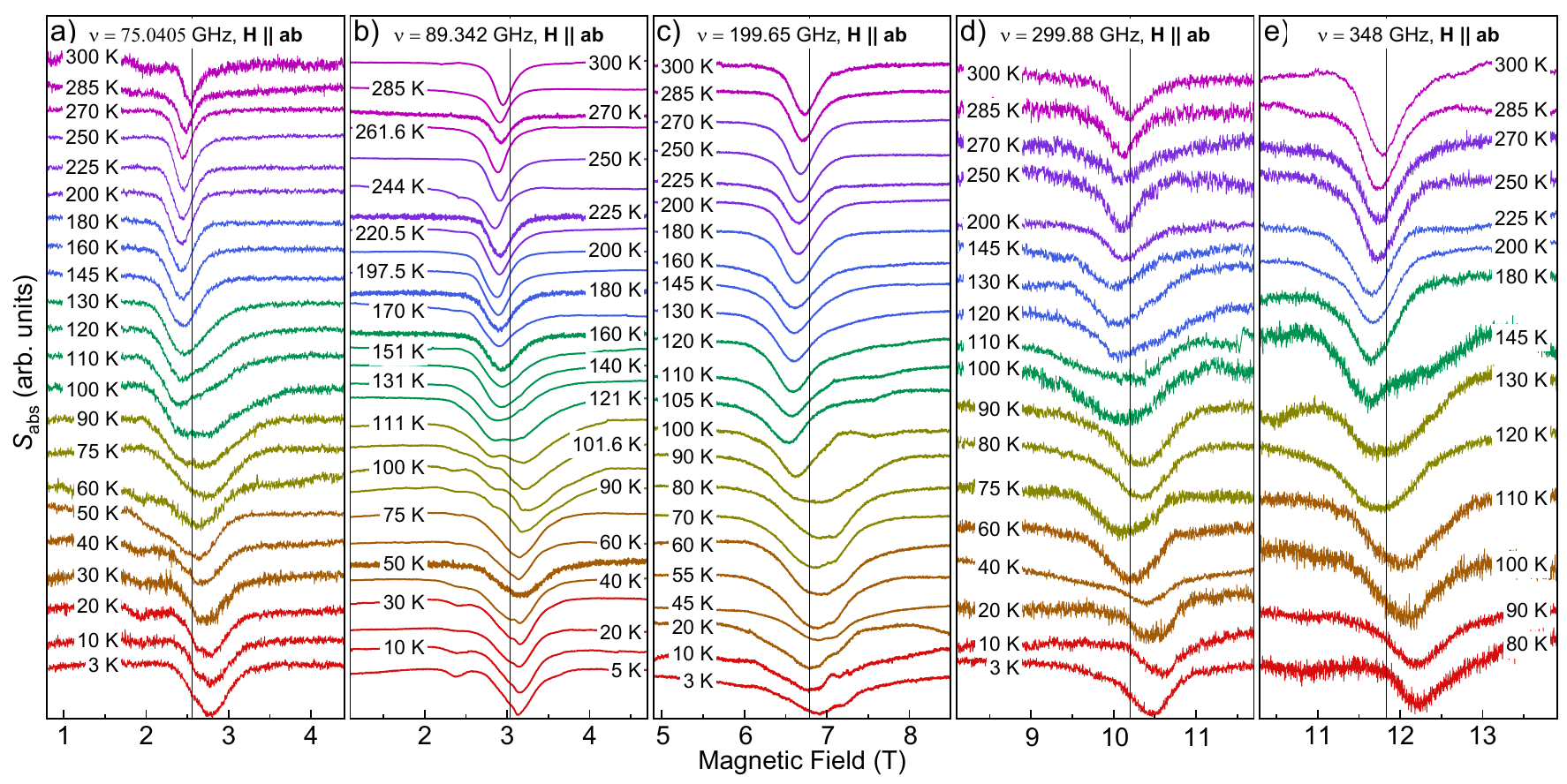}
	\caption{Temperature dependence of the HF-ESR spectra in the \HIIab configuration at fixed excitation frequencies $\nu = $ 75.0405\,GHz (a), 89.342\,GHz (b), 199.65\,GHz (c), 299.88\,GHz (d), and 348\,GHz (e). The spectra are normalized to unity and shifted vertically for clarity. Vertical solid lines represent the expected resonance position of the paramagnetic response.}
	\label{fig:Fig_Spectra_HIIab}
\end{figure*}
\clearpage

\begin{figure}
	\centering
	\includegraphics[width=0.5\linewidth]{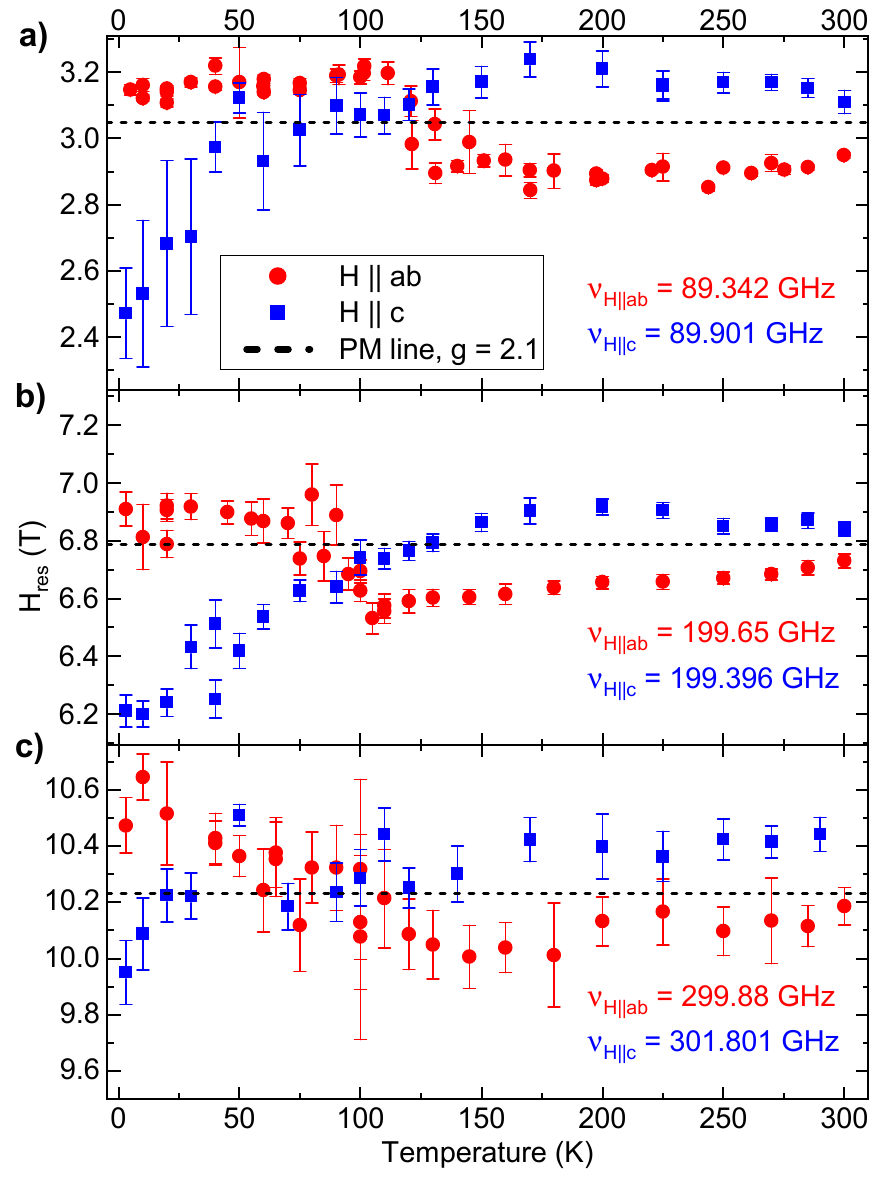}
	\caption{Resonance fields as a function of temperature measured in \HIIab and \HIIc configurations at $\nu \sim 90$\,GHz (a), $\nu \sim 200$\,GHz (b), $\nu \sim 300$\,GHz (c). Horizontal dashed lines represent the expected resonance position of the paramagnetic response at $\Hres^{para}$ according to Eq.~(\ref{eq:Hnu}) with $\text{g} = 2.1$.}
	\label{fig:Fig3}
\end{figure}


\subsection{Magnetic anisotropy}
\label{app:IP}

\begin{table*}[h!]
	\centering
	\caption{Total magnetic anisotropy field \Ha and g-factor values at different temperatures obtained from the fits of the \FvsH dependences measured in the \HIIc and \HIIc configurations. The sign of \Ha defines the type of the anisotropy: It is positive for the easy-plane and is negative for the easy-axis anisotropy, respectively.}
	\label{tab:table1}
	\begin{tabular}{ | r | r | r | c | r | r | r | }
		\hline
		T & $\mu_0 \text{\Ha}$ for \HIIc & $\mu_0 \text{\Ha}$ for \HIIab & average $\mu_0 \text{\Ha}$ & g for \HIIc & g for \HIIab & average g \\
		\hline
		3\,K & $-0.707 \pm 0.118$\,T & $-0.129 \pm 0.109$\,T & \ & 2.045 $\pm$ 0.032 & 2.073 $\pm$ 0.015 & 2.06 $\pm$ 0.01\\
		\hline
		70\,K & $-0.128 \pm 0.084$\,T & $0 \pm 0.158$\,T & $\sim -0.06$\,T & 2.119 $\pm$ 0.025 & 2.085 $\pm$ 0.024 & 2.10 $\pm$ 0.01 \\
		\hline
		270\,K & $0.158 \pm 0.050$\,T & $0.336 \pm 0.090$\,T & $\sim +0.20$\,T & 2.117 $\pm$ 0.015 & 2.092 $\pm$ 0.013 & 2.10 $\pm$ 0.01 \\
		\hline
		300\,K & $0.076 \pm 0.023$\,T & $0.179 \pm 0.052$\,T & $\sim +0.12$\,T & 2.095 $\pm$ 0.007 & 2.099 $\pm$ 0.008 & 2.10 $\pm$ 0.01 \\
		\hline
	\end{tabular}
\end{table*}

\begin{figure*}[ht!]
	\centering
	\includegraphics[width=1.0\textwidth]{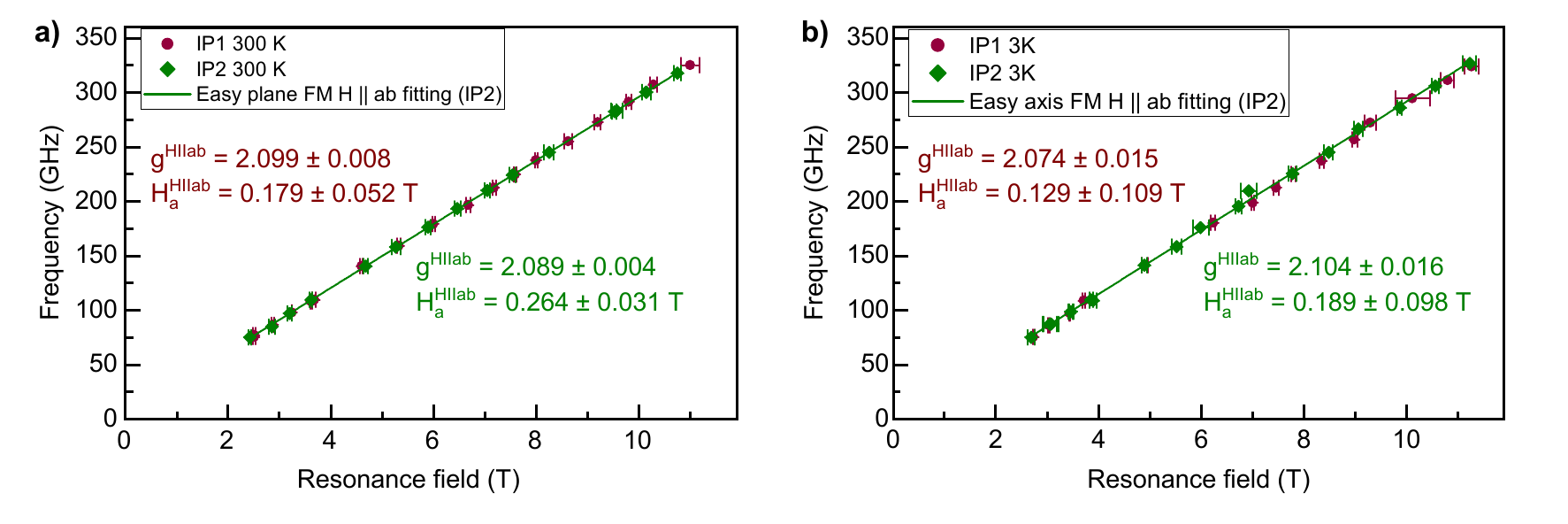}
	\caption{The frequency dependence of the resonance fields $\nu(\Hres)$ at (a) 300 K and at (b) 3 K  for two mutually perpendicular in-plane directions, IP1 (\HIIab) and IP2 (new). Green line indicates the corresponding fitting of the IP2 data. The fitting parameters are written in the inset of the corresponding figures.}
	\label{fig:IP_aniso}
\end{figure*}

To investigate the anisotropy within the $ab$-plane, the \FFGT crystal was rotated by 90 degrees with respect to the original in-plane configuration (IP1). In this second in-plane configuration (IP2) frequency dependences of the resonance fields \FvsH was measured both at $T = 3$\,K, in the magnetically ordered state, and at $T = 300$\,K, $\sim 30$\,K above \TC. The comparison of the results of two configurations IP1 and IP2 is shown in Fig.~\ref{fig:IP_aniso}. As can be seen, at both temperatures, two magnetic field configurations yield the same, within error bars, result, which enables a conclusion, that there is no detectable in-plane anisotropy present in this \FFGT crystal.

\subsection{Magnetization}
\label{app:Magn}

\begin{figure}[ht!]
	\centering
	\includegraphics[width=0.49\linewidth]{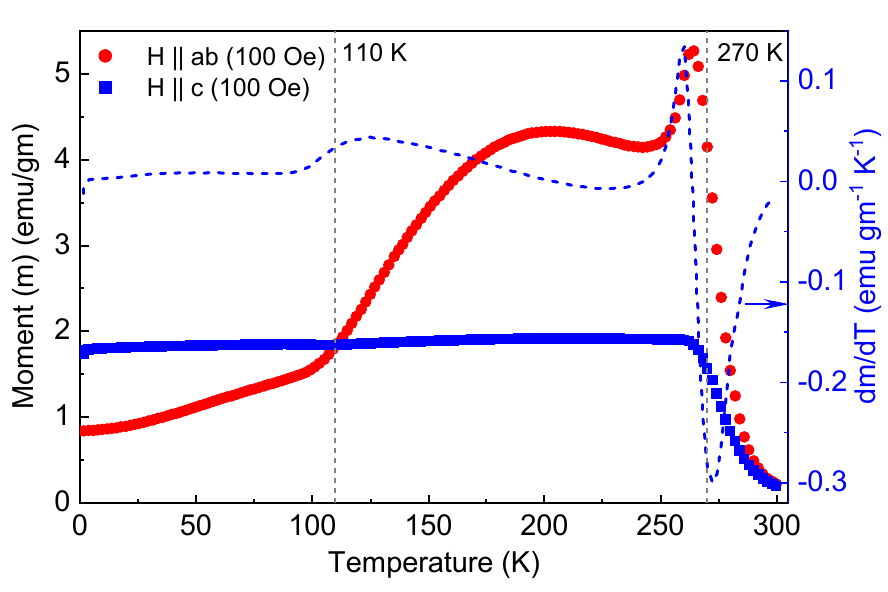}
	\includegraphics[width=0.49\linewidth]{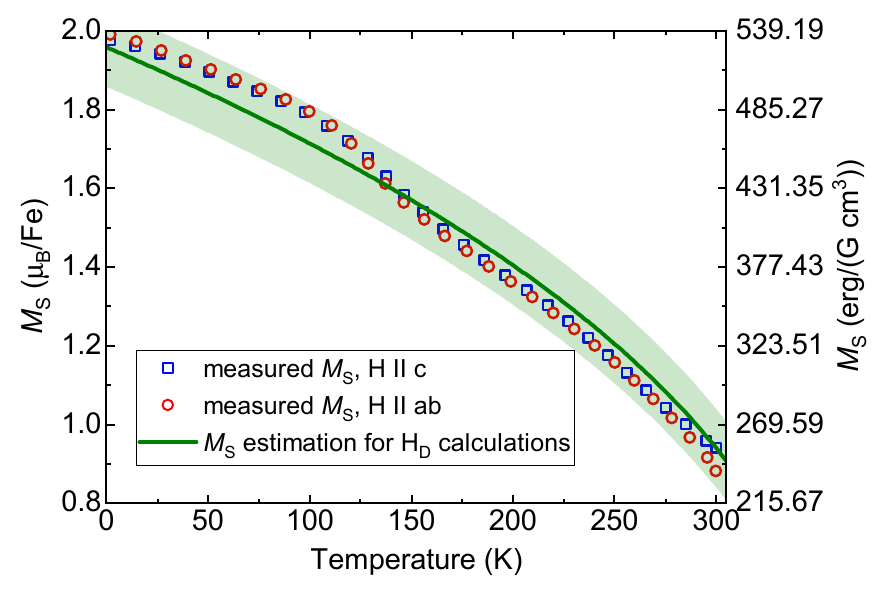}
	\caption{left) Temperature dependence of the magnetization and its derivative measured at the magnetic field of 100\,Oe in \HIIc and \HIIab configurations. right) Temperature dependence of the saturation magnetization $\Msat$ measured in \HIIc (squares) and \HIIab (circles) configurations, the data is taken from \cite{Pal_arxiv2023}. Solid line represents the average $\Msat$ used for the shape anisotropy field ($\HD$) calculation, and the shaded area around this line depicts the uncertainty of the estimated $\Msat$.}
	\label{fig:Fig_Msat}
\end{figure}

\FloatBarrier

\end{document}